\documentclass[prd,amsmath,floats,amssymb,preprint, floatfix,superscriptaddress,nofootinbib,onecolumn]{revtex4} 

\usepackage[plainpages=false, colorlinks=true, anchorcolor=blue, linkcolor=blue, citecolor=blue, bookmarks=false]{hyperref}
\usepackage[T1]{fontenc}
\usepackage{graphicx}
\usepackage{dcolumn}
\usepackage{bm}
\usepackage{longtable}
\usepackage{adjustbox}

\usepackage{float}
\usepackage[caption=false]{subfig}
\usepackage[paperwidth=210mm,paperheight=297mm,centering,hmargin=2cm,vmargin=2.5cm]{geometry}
\usepackage{appendix}

\begin{document}
\include{notations}
\preprint{APS/123-QED}

\title{Search for GeV gamma-ray emission from SPT-CL J2012-5649 with six years of  DAMPE data}

\author{Siddhant Manna}
 \altaffiliation{Email:ph22resch11006@iith.ac.in}
\author{Shantanu Desai}
 \altaffiliation{Email:shntn05@gmail.com}
\affiliation{
 Department of Physics, IIT Hyderabad Kandi, Telangana 502284,  India}





\begin{abstract}
We search for gamma-ray emission from the galaxy cluster SPT-CL J2012-5649 in the energy range from 3 GeV to 1 TeV using the DArk Matter Particle Explorer (DAMPE) telescope. For our analysis, we use three different templates:  point source, radial disk, and radial Gaussian. We do not detect a signal with significance $>3\sigma$ for any of these templates at any location within $R_{200}$ of the cluster center. We obtain 95\% C.L. upper limit on the energy flux ranging between $\sim 10^{-6}$ and $10^{-4} \rm{MeV~cm^{-2}~s^{-1}}$ depending on the energy range. These upper limits are consistent with  a previously reported non-zero flux detected by Fermi-LAT at $6\sigma$ significance. This work represents the first proof of principle search for gamma-ray emission from a single galaxy cluster using DAMPE data.
\end{abstract}


\maketitle
\section{\label{sec:level1}Introduction\protect}

Galaxy clusters are the universe's most massive and stable structures held together by gravity and can be used to probe cosmology~\cite{White1978,Kravtsov2012,Allen2011,Vikhlinin2014} and fundamental physics~\cite{Desai2018,Bora2021,Bora2021b,Bora2022}. Approximately, 80\% of the mass in these clusters is attributed to dark matter, while the intracluster medium (ICM) consisting of hot diffuse gas, accounts for about 10-15\% of their total mass~\cite{Murase2013,Condorelli2023}.  Although galaxy clusters have been mostly detected at all wavelengths from radio waves to X-rays, an unequivocal detection of gamma rays is still an open question. To resolve this issue, in a recent work~\cite{Manna2024}, we carried out a systematic search for gamma-rays from 300 galaxy clusters selected from the 2500 sq. degree SPT-SZ survey~\cite{Bleem15,Bocquet19} based on their values $M_{500}/z^2$ using 15 years of Fermi-LAT data.  From this search, 
we were able to detect   a gamma-ray signature from one  galaxy cluster (viz SPT-CL J2012-5649) with a    significance of  $6.1\sigma$~\cite{Manna2024}. The signal was detected in the energy range between 1-10 GeV with a spectral index of around -3.6. However, given the  point spread function  of Fermi-LAT, we could not ascertain if this signal is due to the intracluster medium or from radio galaxies in the vicinity of the cluster. SPT-CL J2012-5649 (also known as Abell 3667) is a merging cluster with   $M_{500} \approx  5 \times 10^{14} M_{\odot}$  at a redshift of 0.06, and is known to host radio relics~\cite{Omiya,Meerkat}. 
Given the intriguing result, it behooves us to confirm this signal with other operating  gamma-ray detectors, which are sensitive in this energy range. Although we could not see a signal from this cluster at MeV energies~\cite{Manna2024b},  no recent search has been done at GeV energies from this cluster using any other detector. The only other gamma-ray telescope with sensitivity between 1-10 GeV, whose data is publicly available is the DArk Matter Particle Explorer (DAMPE), which was launched in 2015~\cite{Chang2017}.  DAMPE is mainly designed to to study cosmic rays and gamma rays with energies up to 10 TeV~\cite{Chang2017,Ambrosi2019,Chang2014}. Its main science goals are to understand the origin of particle acceleration, cosmic ray propagation, and probing the nature of dark matter~\cite{Chang2017,DAMPE}. Some of the scientific highlights from the DAMPE telescope are reviewed in ~\cite{Fusco20}. As of 2022, DAMPE has detected gamma-ray emission from over 200 sources with significance $>4.5\sigma$~\cite{DAMPE22,Alemanno}.

Although a number of searches have been carried out  with DAMPE for dark matter annihilation from the galactic halo and galactic center~\cite{Alemanno2021,Liu2022,Cheng2023}, there has only been one search for gamma-ray emission from galaxy clusters~\cite{Fan2024}.
In a aforementioned work, a search for  dark matter annihilation using  a sample of 13 clusters from the extended HIFLUGCS catalog was done based on a stacking analysis, to complement a similar search done with Fermi-LAT~\cite{Fan2024}. A signal consistent with Fermi-LAT below 34 GeV was detected and upper limits from 34 GeV to 130 GeV were set (cf. Fig 7 in ~\cite{Fan2024}).

Here, we search for gamma-ray emission from SPT-CL J2012-5649, which is spatially coincident with Abell 3667 in the energy range from 3 GeV to 1 TeV using six years of DAMPE data. This manuscript is structured as follows. The data analysis setup is discussed in Sect.~\ref{sec:da}. Our results are described in Sect.~\ref{sec:results}. We conclude in Sect.~\ref{sec:conclusions}



\section{Data Analysis}
\label{sec:da}
\label{sec:level2}
The DAMPE telescope can detect  cosmic rays and gamma rays from  3 GeV - 10 TeV~\cite{Chang2017,Ambrosi2019,Chang2014}. It has an angular resolution of $\leq 0.2^{\circ}$ at 100 GeV  with a field of view of about 1.0 sr. DAMPE has  a much finer energy resolution ($\sim$ 1.0\% at 100 GeV) compared to Fermi-LAT across a broad range of energies~\cite{Chang2017}. However its acceptance is seven times smaller than Fermi-LAT. More technical details about the DAMPE detector, including its on-orbit performance calibrations are reviewed elsewhere~\cite{Chang2017,Chang2014,Ambrosi2019}.

We conducted a gamma-ray search for  SPT-CL J2012-5649  (R.A.= 303.11$^{\circ}$, Dec.= -56.83$^{\circ}$) utilizing six years of DAMPE data, covering the time range from  from January 1, 2016 to January 1, 2022 (\texttt{MET 94608000 - 283996800}). 
The data was selected within a region of interest (ROI) with a radius of $10^{\circ}$ centered on the cluster center, as determined by SPT observations. The energy range for the analysis was set between 3 GeV and 1000 GeV (1 TeV). We used \texttt{DAMPE Science Tools (DmpST)}~\cite{Duan} software for our analysis. This software uses the  photon event data, spacecraft files, Monte Carlo (MC) instrument response functions (IRFs), and a predefined model of gamma-ray sources to perform the analyses. We filtered the photon events based on the direction, time, energy, and the trigger type. We selected only the events that met the criteria for the High-Energy Trigger (HET). Data gathered during the South Atlantic Anomaly or during periods of intense solar flares were excluded from our analysis. We also calculated the angles between the target source and the $Z$-axis of the spacecraft. If the angle is  larger than $60^{\circ}$, the target source is not in the field of view of \texttt{DAMPE} and we do not include that data. To visualize the photon information about energy, time, and location, we consider the photon count maps and event maps. We then calculate the total exposure using the  livetime cube and proceed to define an input model file, which includes the target source and diffuse emission sources. The observed livetime for our analysis is around $1.43 \times 10^{8}$ seconds. 
For our analysis, we considered three different search templates: a point-source template, extended radial disk, and radial Gaussian templates. Each of these templates also needs  a power-law spectral model to characterize the gamma-ray emission from the source. The power-law spectral index was set to $\Gamma=-2$, consistent with previous studies~\cite{Baghmanyan2022,Manna2024}. We used  $0.1^{\circ}$ pixel resolution for spatial binning and ten energy bins spanning the  range between 3-1000 GeV.
During the fitting process, we kept the normalization of the source model as a free parameter to allow for optimal adjustment. To account for the diffuse gamma-ray emission in the region, we utilized the Galactic diffuse emission model \texttt{(gll\_iem\_v07.fits)} along with an isotropic component, which we modelled using a Power-Law spectral model.  We allowed the normalizations of the Galactic foreground and isotropic diffuse emission templates to vary freely during the fitting process. We performed a standard binned likelihood analysis to determine the best-fit model parameters for the source's spectral and positional characteristics.

 
\section{Results}
\label{sec:results}
\begin{figure}
\hfill
\centering
\begin{adjustbox}{right=0.7\columnwidth}
\includegraphics[width=0.9\columnwidth]{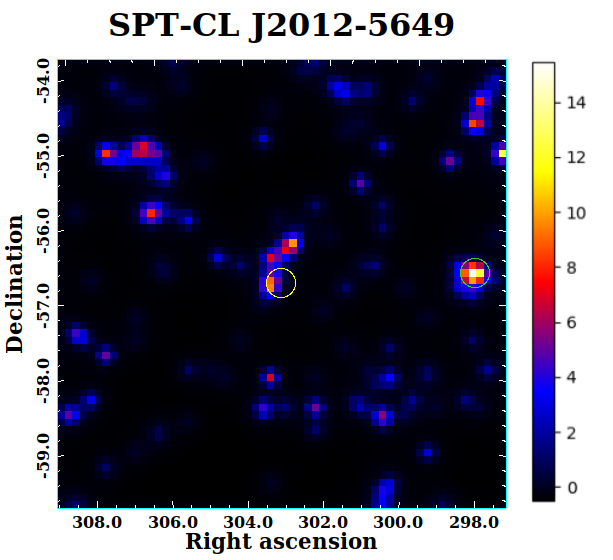}
\end{adjustbox}
\caption{TS map of the SPT-CL J2012-5649 cluster (left) and TS map scale (right) generated using \texttt{DmpST} in the energy band $3- 1000$ GeV. We used $0.1^{\circ}$ pixel resolution for the spatial binning and a point source template. The SPT-SZ position of the cluster is shown at the center of the map enclosed by a circle of radius $0.2^{\circ}$ in yellow which corresponds to its $R_{200}$ at a redshift of 0.06. We also show the new source by a circle of radius $0.2^{\circ}$ at (RA, Dec) = (298.4, -56.6) from which we detect emission of 3.94~$\sigma$ in green.}
\label{fig:figure1}
\end{figure}

\begin{figure}
\hfill
\centering
\begin{adjustbox}{right=0.7\columnwidth}
\includegraphics[width=0.9\columnwidth]{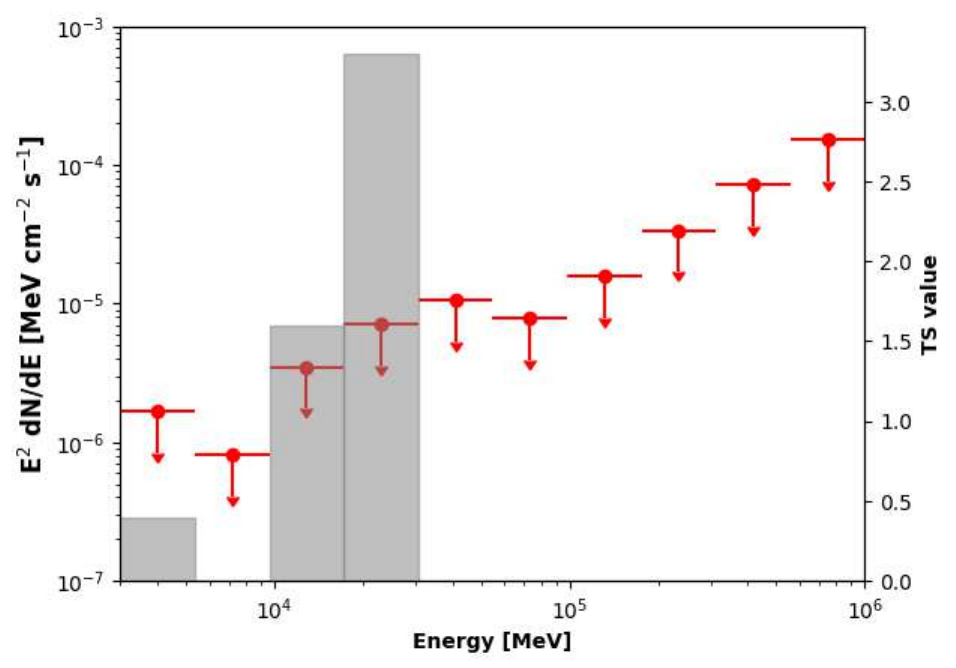}
\end{adjustbox}
\caption{The SED plot for the SPT-CL J2012-5649 galaxy cluster using a point source template and 10 energy bins is shown. We have shown the upper limits in red color. The energy bins in grey bars have positive TS values , 0 < TS < 9. On the right side of the plot, the TS scale is provided to indicate the detection significance. The TS values for the three bins in grey are equal to 0.4, 1.6, and 3.3 respectively.
The results from radial disk and radial Gaussian templates are almost the same as the point source template. }
\label{fig:figure2}
\end{figure}

\begin{figure}
\hfill
\centering
\begin{adjustbox}{right=0.7\columnwidth}
\includegraphics[width=0.9\columnwidth]{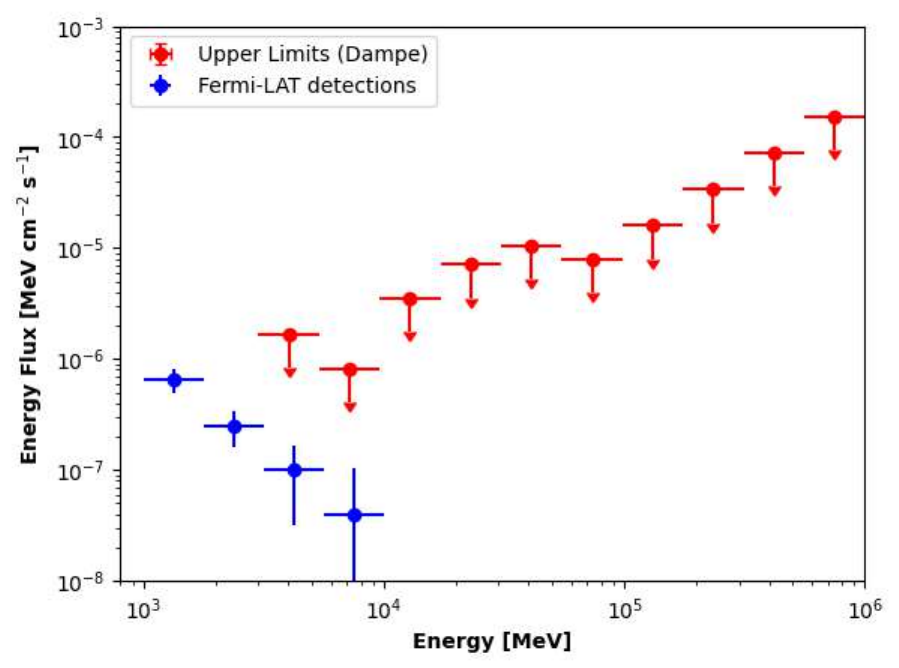}
\end{adjustbox}
\caption{The SED for SPT-CL J2012-5649 using a point source template for the energy range 1-1000 GeV. Blue data points represent the measured energy spectrum obtained from Fermi-LAT data~\cite{Manna2024}. Red points show the 95\% C.L. upper limits derived from DAMPE data. }
\label{fig:figure3}
\end{figure}

A   comparison of the best-fit central positions, radial extension and natural log-likelihood values for point, radial disk, and radial Gaussian spatial models can be found in Table~\ref{tab: Table I}.
The  Test Statistic (TS) map for the SPT-CL J2012-5649 cluster can be found in Fig.~\ref{fig:figure1} for a point source template. TS is defined in the same way as for Fermi-LAT and IceCube analysis~\cite{Manna2024,OJ287,Pasumarti}. For the null hypothesis, the probability distribution of TS behaves like a $\chi^2$ distribution with degrees of freedom determined by the difference in the number of free parameters for the signal model compared to the background model~\cite{Mattox96}. For one additional degree of freedom,  the statistical significance of any possible detection   can  be approximately  obtained from the square root of the TS value.
At the cluster center, the point source template yielded a maximum TS value of 5.1, corresponding to a significance of 2.3$\sigma$. Using the radial disk and radial Gaussian templates, TS values of 4.5 and 4.0 were found at the cluster center, respectively. A maximum TS value of 15.5 was found at coordinates (RA, Dec) = (298.4$^{\circ}$, -56.6$^{\circ}$) using the point source template, corresponding to a statistical significance of 3.94$\sigma$. For comparison, the radial disk and radial Gaussian templates yielded maximum TS values of 14.6 and 13.8, respectively, at the same coordinates. This position of the  maximum TS value is  at an offset of 2.59$^\circ$ with respect to  the SZ cluster center. This is much larger than the angular size of the cluster ($\theta_{200}= 0.19^{\circ}$). 
In Figure~\ref{fig:figure2}, we plot the Spectral Energy Distribution (SED)  for the aforementioned cluster using a point source template and ten energy bins. The energy bins with a Test Statistic (TS) value between zero and nine are shown as grey bars, where the height of the bar indicates the significance of the gamma-ray signal. Since none of the bins have TS values greater than nine, we show   upper limits for all the bins, which are depicted by red arrows. The upper limits correspond to 95\% confidence level.   In Figure~\ref{fig:figure3}, we constructed the differential energy spectrum of SPT-CL J2012-5649, we overlay the Fermi-LAT observation for this cluster along with DAMPE. The Fermi-LAT data, as presented in~\cite{Manna2024}, provide the measured energy spectrum, represented by blue data points in Fig.~\ref{fig:figure3}. We incorporated the upper limits from DAMPE data, shown as red points in the same figure. This combined analysis allows for a comprehensive characterization of the gamma-ray emission from this source over a wide energy range of 1-1000 GeV.  The upper limits on the differential energy flux are in the range from $\sim (10^{-6}-10^{-4}) \rm{MeV~cm^{-2}~s^{-1}}$ depending on photon energy (cf. Fig.~\ref{fig:figure3}).
Therefore, we can see that the DAMPE upper limits are consistent with the Fermi-LAT results, given the higher sensitivity of Fermi-LAT.  We shall also present a combined analysis of Fermi-LAT and DAMPE data for the cluster in Appendix A.

\begin{table}[h!]
    \centering
    \caption{\label{tab: Table I} Comparison of the best-fit central positions, radial extensions, and log of the likelihood ($\ln (L)$) values for point, radial disk and radial Gaussian spatial models.}
    \begin{tabular}{|l|c|c|c|c|}
        \hline
        \textbf{Template} & \multicolumn{2}{c|}{\textbf{Central Position}} & \textbf{Radial Extension} & $\mathbf{\ln(L)}$ \\ \cline{2-3}
        & \textbf{RA} & \textbf{DEC} & & \\ \hline
        \textbf{Point} & 303.113  & -56.830  & N/A & \text{7.27} \\ \hline
        \textbf{Radial Disk} & 303.113  & -56.830 & Radius = 0.2$^\circ$  & \text{7.58} \\ \hline
        \textbf{Radial Gaussian} & 303.113  & -56.830  & Sigma = 0.2$^\circ$ & \text{7.85} \\ \hline
    \end{tabular}
\end{table}

\hfill
\hfill
\hfill
\clearpage
\section{\label{sec:conclusions}Conclusions\protect}
In a recent work~\cite{Manna2024}, we had detected gamma-ray emission from SPT-CL J2012-5649 using 15 years of Fermi-LAT data in the energy range between 1-10 GeV with a significance of  6.1$\sigma$. In this work, we carry out a similar search for high energy emission in the same energy range from this cluster using data from another existing  gamma-ray telescope, viz DAMPE. For our analysis, we use six years of publicly available data from the DAMPE telescope and used three different templates for our analysis: point source, radial disk, and radial Gaussian, similar to our Fermi-LAT analysis. We do not find statistically significant gamma-ray emission ($>3\sigma$) at the cluster center or within its $R_{200}$. Consequently, we report upper limits in all the energy bins. Our 95\% C.L. upper limits for SPT-CL J2012-5649 in the energy range from 3 GeV-1 TeV are shown in Fig.~\ref{fig:figure3} along with the corresponding results from Fermi-LAT.
The upper limits on the differential energy flux, which we obtain above 10 GeV are between $\sim 10^{-6}-10^{-4} \rm{MeV~cm^{-2}~s^{-1}}$.

Therefore, this is the first proof of principle analysis of gamma-ray searches using a single galaxy cluster  from DAMPE.  The next data release consisting of seven years of data will also happen soon (Zhaoqiang Shen, private communication), which will enable us to easily extend this analysis with the new data.
in the near future more detailed gamma-ray analysis of this cluster should be possible using the upcoming Very Large Area Gamma-ray Space Telescope (VLAST)~\cite{VLAST}, which has been proposed by the DAMPE team.

\begin{acknowledgments}

We acknowledge data resources from DArk Matter Particle Explorer (DAMPE) satellite mission supported by Strategic Priority Program on Space Science, and data service provided by National Space Science Data Center of China. We are thankful to Zhaoqiang Shen for pointing us to the DAMPE data and software.

\end{acknowledgments}

\bibliography{references}
\appendix
\section{Combined Analysis of Fermi-LAT and DAMPE}
\begin{figure}
\hfill
\centering
\begin{adjustbox}{right=0.7\columnwidth}
\includegraphics[width=0.9\columnwidth]{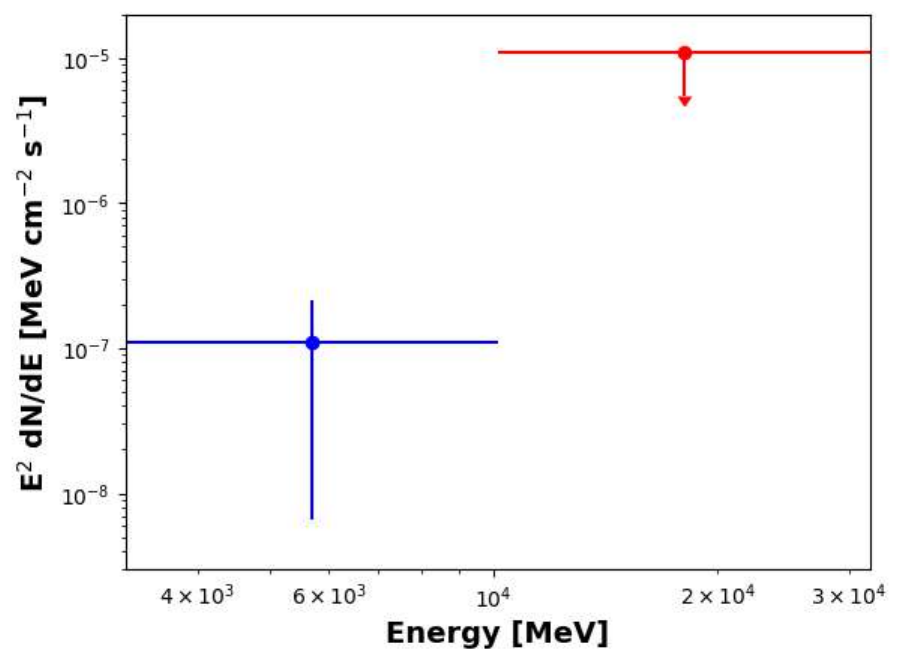}
\end{adjustbox}
\caption{ Results from  the combined analysis of Fermi-LAT and DAMPE data for SPT-CL J2012-5649, focusing on the energy range from 3.2 to 32 GeV. Using two energy bins, we found a significant detection (TS $> 9$) in the first bin, which is represented in blue. The SED  for the first bin was calculated using the weighted mean of Fermi-LAT and DAMPE SED  according to Eq.~\ref{eq:comb}. For the second bin, we calculated upper limits (TS $< 9$), shown in red.}
\label{fig:figure4}
\end{figure}

We performed a combined analysis of Fermi-LAT and DAMPE data for SPT-CL J2012-5649. We collected Fermi-LAT data from April 5, 2008 to September 1, 2024 \texttt(MET 239587201-746841605) and DAMPE data from January 1, 2016 to January 1, 2022  (MET 95608000 - 283996800). We focused on a specific region of the sky, viz. a $10^{\circ}$ circle around the cluster center. We only considered particles with energies between 3.2 GeV and 32 GeV and used $0.1^{\circ}$-pixel resolution for spatial binning and two energy bins (3.2-10 GeV and 10-32 GeV). We added  the TS values from both the  telescopes to get the combined results. The combined SED plot can be found in  Figure~\ref{fig:figure4}. 
For the first energy bin we get TS $>9$ (TS = 9.8), whereas for the other bin we get TS $<9$ (TS = 6.6). Therefore, for the first energy bin, we combined the  SEDs by calculating the weighted mean after adding the SEDs from each detector,  weighted by their uncertainties  as follows:
\begin{equation}
F_{comb} =  \frac{ \sum \limits_{i=1}^2 F_i/\sigma_i^2}{ \sum \limits_{i=1}^2  1/\sigma_i^2} 
\label{eq:comb}
\end{equation}
where $F_i$ is the SED  in the individual detector with uncertainty $\sigma_i$ and $F_{comb}$ is the combined SED. The reciprocal of the uncertainty in $F_{comb}$ is  given by:
\begin{equation}
\sigma_{F} = \frac{1}{\sqrt{\sum \limits_{i=1}^N 1/\sigma_i^2}}.
\label{eq:errcomb}
\end{equation}
This combined flux is shown in  the blue data point in Fig.~\ref{fig:figure4}. Since the TS values of both the detectors as well as the combined 
TS value are less than nine for the second energy bin, we show the smaller of the two flux limits in Fig~\ref{fig:figure4}, which is shown using the red downward arrow.

\end{document}